\documentclass[12pt]{article}
\usepackage{graphicx}
\usepackage{graphicx}
\begin{document}

\title{Thermal conductivity of \\ superconducting $\rm MgB_2$}
\author{
E. Bauer, Ch. Paul, St. Berger, S. Majumdar, and H. Michor \\
Institut f\"ur Experimentalphysik \\
Technische Universit\"at Wien, A-1040 Wien, Austria \\
M. Giovannini and A. Saccone \\
Dipartimento di Chimica e Chimica Industriale,\\
 Universit\`a di Genova, I-16146 Genova, Italy \\
A. Bianconi \\
Unita INFM and Dipartimento di Fisica, \\
Universit\`a di Roma "La Sapienza", I-00185 Roma, Italy
}
\maketitle

\begin{abstract}

Thermal conductivity of superconducting $\rm MgB_2$
was studied in both the superconducting and the normal state
region. The latter is almost equally determined by 
the electronic -  and the lattice  
contribution to the total thermal conductivity.
In the superconducting state, however, the lattice 
contribution is larger. The electronic thermal 
conductivity below $T_c$ was derived from the experimental
data considering the Bardeen-Rickayzen-Tewordt theory
together with the model of Geilikman. The analysis shows
that electron scattering on static imperfections 
dominates.

\textbf{key words}: $\rm MgB_2$ superconductivity,
thermal conductivity, specific heat

\end{abstract}


\section{Introduction}

Superconductivity with a remarkably high transition temperature
$T_c \approx 39$~K was recently discovered in MgB$_2$
\cite{akimitsu}. The subsequent investigation of the boron isotope 
effect by Bud'ko {\sl et al.}~\cite{budko} revealed a partial isotope 
exponent $\alpha_{\rm B}\approx 0.26$ (corresponding to $\Delta T_c=1$\,K)
which appears to be consistent with a phonon-mediated BCS 
superconducting (SC) mechanism. Also other experimental reports as e.g.\
on specific heat~\cite{kremer,waelti} argued that their data 
can be accounted for by a conventional, s-wave type BCS-model.
Investigations of the SC gap
of $\rm MgB_2$ by means of Raman scattering are also consistent with 
an isotropic s-wave gap with a moderate coupling
$2 \Delta \approx 4.1 k_BT_c$ \cite{Chen}.   
However, a theoretical analysis of the temperature dependence
of the upper critical field $H_{c2}(T)$ in terms of Eliashberg
type models by Shulga {\sl et al.}~\cite{shulga} demonstrates that
the shape and the magnitude of the upper critical field of 
MgB$_2$ can definitely not be accounted for by an isotropic 
single band model, but may successfully be described 
within a multi-band Eliashberg model with various options. 
A careful calorimetric investigation of the SC
parameters of MgB$_2$ by Wang {\sl et al.}~\cite{wang} gave
even more direct evidence against the arguments for simple
isotropic BCS type superconductivity: 
``The nearly quadratic dependence of $C(T)$ versus $T$ at 
$T \ll T_c$, its non-linear field dependence, and the discrepancy 
between the electron-phonon coupling constant $\lambda_{ep}$ as 
determined by the renormalization of the electron density of states 
($\lambda_{ep}\sim 0.6$) and by McMillan's equation for isotropic 
superconductors ($\lambda_{ep}\sim 1.1$), are inconsistent with a 
single isotropic gap''. 
Direct hints for a non BCS temperature dependence of the gap energy
$\Delta (T)$ were also obtained by tunneling experiments on
MgB$_2$/Ag and MgB$_2$/In junctions~\cite{plecenik}.

Thermal conductivity $\lambda$ is one of those transport
coefficients  which exhibits non-zero values 
in both the normal and the 
SC state. The temperature dependence of
$\lambda$ allows to distinguish between the most important 
interactions present in a superconductor. In particular,
the interaction of electrons with phonons
are recorded in the magnitude of $\lambda(T)$. Moreover,
scattering of these particles by static imperfections like
impurities, defects or  grain boundaries are reflected. 

The aim of the present work is to derive the temperature
dependent thermal conductivity of $\rm MgB_2$ and to analyse
the data with respect to the electronic - and the lattice
thermal conductivity both in the 
normal and the SC state. Moreover, we present
resistivity and specific heat measurements 
in order to characterise the investigated sample.   

\section{Experimental}

A $\rm MgB_2$ sample of about 1.3 g was 
synthesized by direct reaction of the
elements. The starting materials were 
elemental magnesium (rod 99.9 mass \%
nominal purity) and boron (99.5 \% powder, crystalline, 
$<$ 60 mesh, 99.5 mass \%). 
The elements in a stoichiometric ratio were enclosed in a cylindric
tantalum crucible sealed by arc welding under argon atmosphere. The
tantalum crucible was then sealed 
in an iron cylinder and heated for one hour
at $800^{\circ}$~C and two hours at $950^{\circ}$~C in a furnace. 
The sample characterized by x-ray 
diffraction show pure $\rm MgB_2$ phase;
only one very weak peak due to extra
phase has been found. 

The thermal conductivity measurement was performed 
in a flow cryostat on a cuboid-shaped sample 
(length: about 1 cm, cross-section: about 2.5~mm$^2$), 
which was kept cold by anchoring one end of the 
sample onto a thick copper panel mounted on the 
heat exchanger of the cryostat. The temperature 
difference along the sample, established by 
electrical heating, was determined by 
means of a differential thermocouple 
(Au + 0.07 \% Fe/Chromel). The measurement was performed 
under high vacuum and 3 shields mounted around 
the sample reduced the heat losses due to radiation 
at finite temperatures. The innermost of these shields 
is kept on the temperature of the sample via an 
extra heater maintained by a second temperature controller.

Resistivity data were taken from bar shaped samples applying
a standard 4-probe d.c. technique at temperatures down to  0.5~K and in 
magnetic fields up to 12~T. 

Specific heat measurements were carried out on a sample 
of about 1~g in the temperature range 5\,K--50\,K 
using a quasi adiabatic step heating technique.

\section{Results and discussion}

\begin{figure}[h]
\includegraphics[width=12cm,height=9cm]{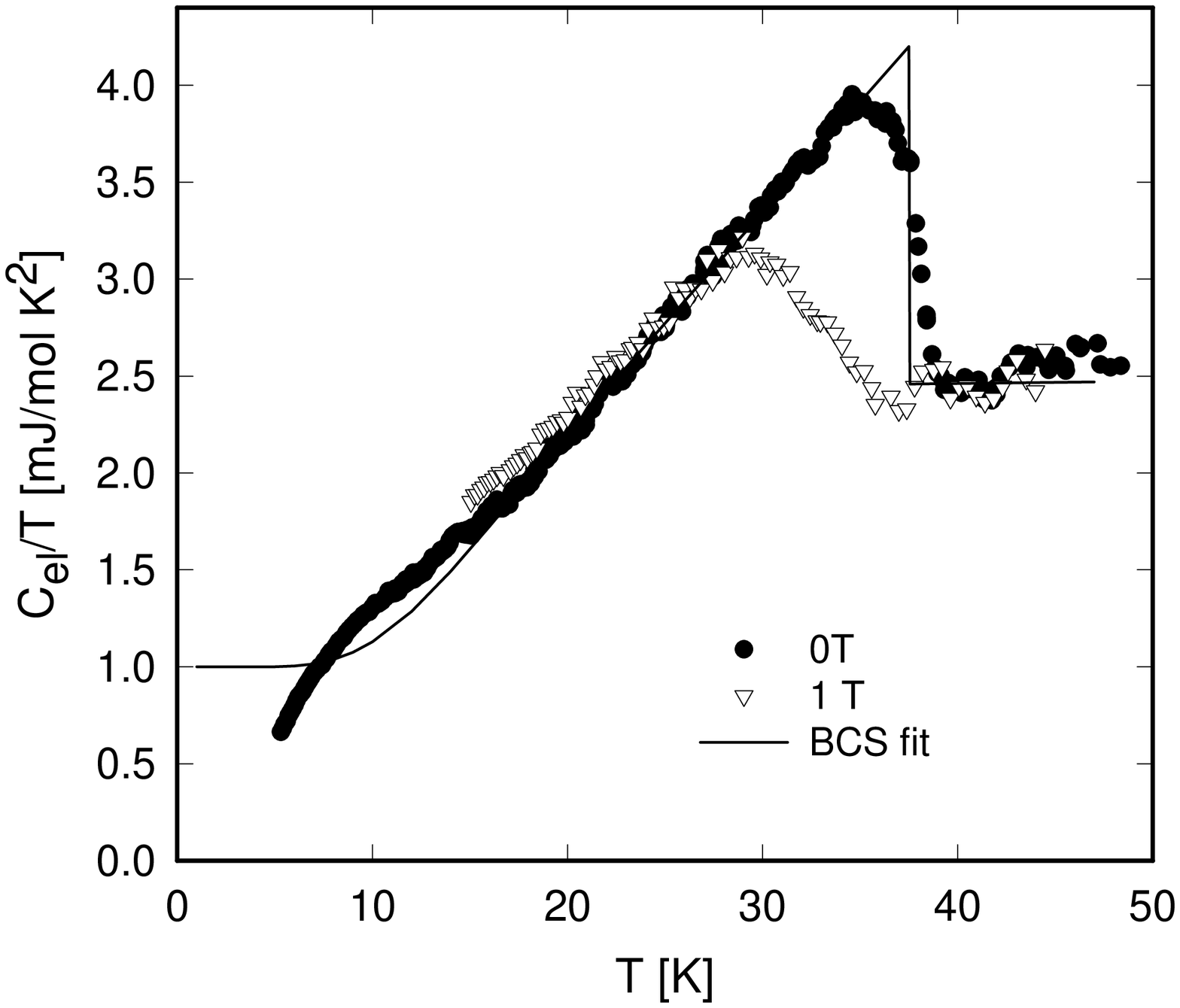}
\caption{The electronic specific heat of $\rm MgB_2,$ $C_{el}(T)/T$ 
versus $T$ obtained by subtracting the lattice contribution,
$C_{lat}=C^{9T}-\gamma T,$ where $\gamma\simeq 2.4$ mJ/mol\,K$^2.$ 
The full line indicates a BCS temperature dependence of $C_{el}$ as 
explained in the text.}
\label{fig0}
\end{figure}

In order to give a direct proof of the SC bulk
properties of our MgB$_2$ sample prior to the 
transport measurements we checked the specific heat .
These specific heat measurements performed in zero-field, 1 and 9~T 
showed reasonable agreement with results previously reported 
 \cite{kremer,waelti,wang}. The thermodynamic
mean transition temperature $\overline{T_c}$ of our sample is 37.5\,K. 
As already noted in the introduction there have been distinctly
different and partly controversial conclusions suggested in Refs.
\cite{kremer,waelti,wang}, although their raw data are in
fair agreement with each other. 
Therefore, we show in figure \ref{fig0} 
the temperature dependence of the electronic specific heat, 
$C_{el}(T)$ versus $T,$ obtained by subtracting the
lattice heat capacity deduced from the 9\,T specific heat data.
Of course, 9\,T are insufficient to obtain a complete suppression
of superconductivity in MgB$_2$, but there is already a large 
reduction of the order parameter combined with a dramatic broadening 
of the transition (see Ref.~\cite{wang} for comparison of 10, 14 and 
15\,T data). Thus, we obtained a Sommerfeld coefficient of the 
normal-state electronic specific heat $\gamma \simeq 2.4$(2) mJ/mol\,K$^2$
by extrapolating the 9\,T data in $C/T$ versus $T^2$ from
30 -- 100 K$^2$ to zero temperature which is already close to
$\gamma = 2.7\pm 0.15$ mJ/mol\,K$^2$ obtained from the 14 and 16\,T
measurement by Wang {\sl et al.}~\cite{wang}.
The important point to emphasize in figure \ref{fig0} is
the non BCS-like temperature dependence of the SC-state electronic
specific heat. In fact, we observed a similar deviation from a simple
BCS temperature dependence (solid line, figure \ref{fig0})
as previously reported by 
Wang {\sl et al.}~\cite{wang}
($C_{el}^{BCS} = 8.5 \gamma T_c \exp[{-0.82\Delta (0)/k_BT}]$ 
for $T < 0.4 T_c$). This discrepancy is supposed to be
indicative for the opening of an additional 
gap below about 10\,K .
The solid line in figure 1 indicates that a fraction of
electrons corresponding to a normal state $\gamma\sim 1.4$ mJ/mol\,K$^2$
is tentatively  accounted for by the BCS-fit with 
$\Delta (0) / k_B T_c \simeq 1.9$ while a second fraction 
corresponding to $\gamma\sim 1.0$ mJ/mol\,K$^2$ contributes to 
the smaller gap opening at temperatures well below $T_c/2.$

\begin{figure}[h]
\includegraphics[width=10cm,height=6.5cm]{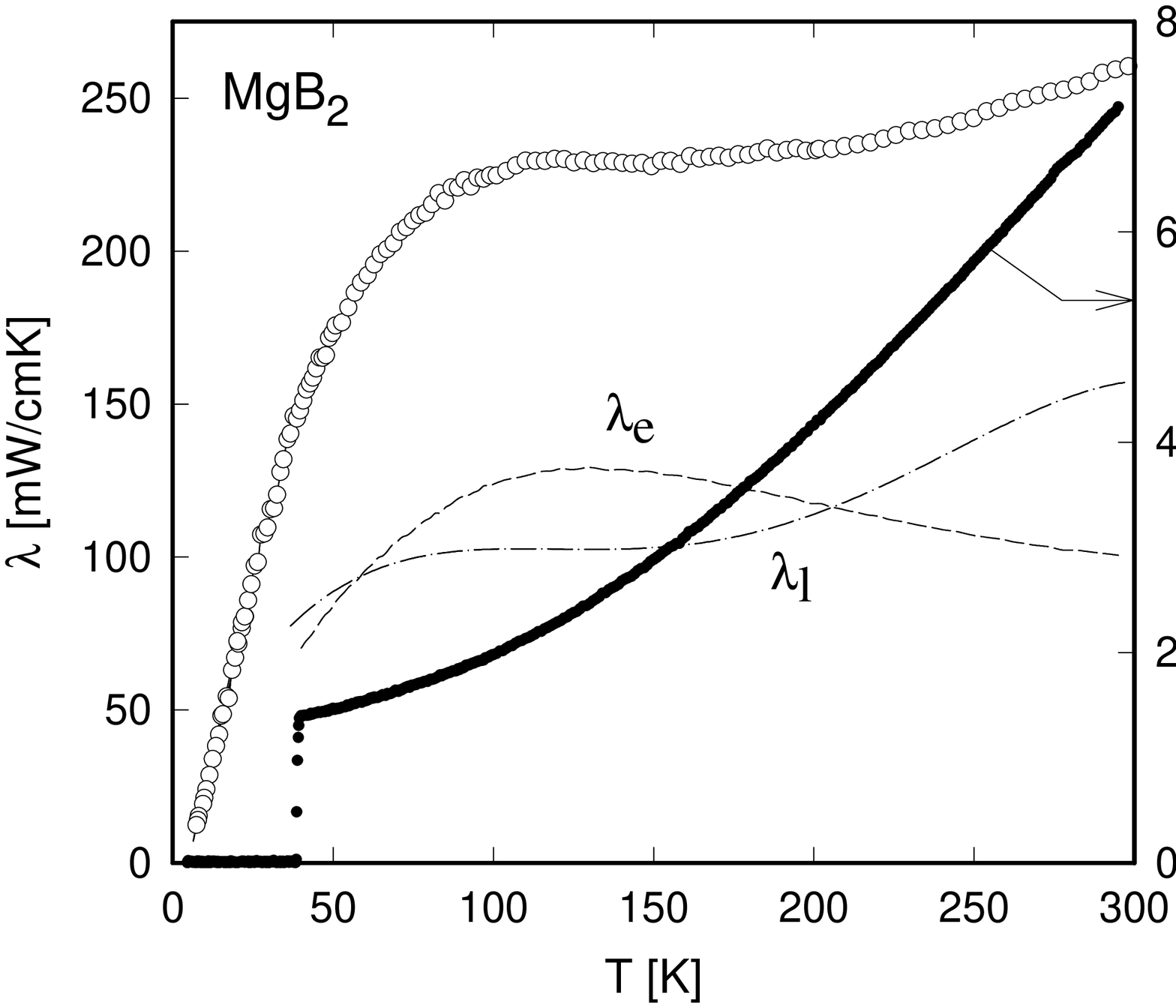}
\caption{Temperature dependent thermal conductivity $\lambda (T)$ (left axis),
and electrical resisitivity $\rho(T)$ (right axis)
of $\rm MgB_2$. The dashed and the dashed-dotted lines are the 
electronic - and the lattice contributions 
$\lambda_e$ and $\lambda_l$, respectively.}
\label{fig1}
\end{figure}

To further screen the quality of the sample, 
temperature and magnetic field dependent
resistivity measurements $\rho(T,H)$ 
were performed from 0.5~K up to room temperature.
Shown in figure \ref{fig1} (right axis) is $\rho(T)$
at zero field. The transition into the SC state
occurs at $T_c = 38.9$~K, which is in fine agreement with already
published data. The RRR ratio of this polycrystalline sample
is about 6.  The resistivity behaviour in the normal state region
matches  a dependence according
to $\rho(T) = \rho_0 + AT^2$ with the residual 
resistivity $\rho_0 = \rm 12.5~\mu \Omega cm$ and the coefficient
$\rm A = 9 \times 10^{-4}~\mu \Omega cm$. A $T^2$ behaviour of
$\rho(T)$ was recently reported  \cite{Jung} and it seems to reflect 
interactions between charge carriers. A study of the Hall coefficient 
implies that electrical transport is dominated 
by holes \cite{Kang}.
The value of the coefficient $A$, however, 
is significantly smaller than that known e.g., for 
highly correlated electron systems, but seems to reflect the 
modest density of states at Fermi energy \cite{Kortus}.
It should be mentioned that other power laws with an  exponent
close to 3 were reported for sintered material of $\rm MgB_2$
\cite{Finnemore,Canfield}. 
Measurements of the resistivity down to 0.5~K 
and in fields up to 12~T indicate that $H_{c2}$ is well above that
limit. Interestingly, the transition region becomes much broader
with increasing fields, but different values of imprinted
currents (from 5 to 40 mA) do not change the width of the 
transition.

The temperature dependent thermal conductivity $\lambda$ of 
$\rm MgB_2$ is shown in figure \ref{fig1}. The overall
behaviour of $\lambda (T)$ is typical of an intermetallic
compound where scattering on static imperfections prohibits
a pronounced maximum to occur, as it is the case in pure 
and simple metals. Moreover, the absolute magnitude appears
to be of that order, as usually found for intermetallics. 
Anomalous behaviour of $\lambda(T)$ in the proximity of $T_c$
is not observed and a local maximum or a pronounced shoulder
below $T_c$ do not occur in the investigated sample.

Generally, the total thermal conductivity of metals consists of 
a sum of 
an electronic contribution $\lambda_e$ and a lattice contribution 
$\lambda_l$:
\begin{equation}
\lambda = \lambda_e + \lambda_l .
\label{eq1}
\end{equation}
In order to separate both contributions from the total
measured effect, the Wiedemann-Franz law is applied,
assumed to be valid, at least, in simple metals. This model
relates the electrical resistivity $\rho$ with 
the electronic contribution to the thermal conductivity
$\lambda_e$ and can be expressed as
\begin{equation}
\lambda_e(T) = \frac{L_0 T}{\rho(T)},
\end{equation}
where $L_0 = 2.45 \times 10^{-8}$~$\rm W \Omega K^{-2}$ is the
Lorenz number.

Using equation \ref{eq1} and taking into account the 
appropriate values of the electrical resistivity in the 
normal state region of $\rm MgB_2$ 
(compare figure \ref{fig1}, right axis)
allows to split $\lambda$ into $\lambda _e$ (dashed line, figure
\ref{fig1}) and $\lambda_l$ (dashed-dotted line, figure
\ref{fig1}). This type of analysis indicates that both
contributions are almost equal in the entire temperature 
range of the normal state region of $\rm MgB_2$. 

According to Matthiessen's rule both $\lambda_e$ and 
$\lambda_l$ are limited owing to various scattering processes,
which can be expressed in terms of a thermal
resistivity $W$. In the case of non-magnetic materials,
the following temperature dependence of the electronic 
contribution to the total measured quantity is assumed to be
valid \cite{Tye}:
\begin{equation}
1/\lambda_e (T) \equiv  W_e (T) = W_{e,0}(T) + W_{e,ph} (T)
= \frac{\alpha}{T} + \beta T^2,
\label{eq2}
\end{equation}
where the subscripts (e,0) and (e,ph) refer to interactions
of the conduction electrons with static imperfections and 
thermally excited phonons, respectively; 
$\alpha$ and $\beta$ are material constants.

\begin{figure}[h]
\includegraphics[width=10cm,height=6.5cm]{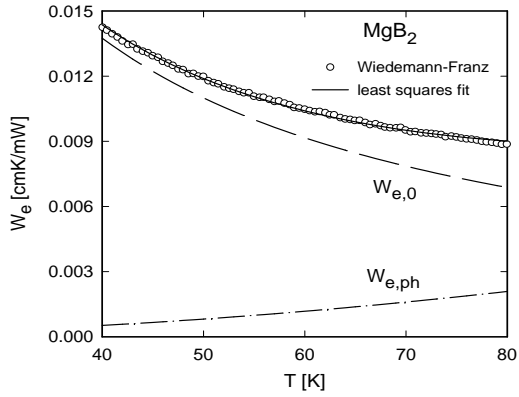}
\caption{Temperature dependent electronic 
thermal resistivity $W_e (T)$ 
of $\rm MgB_2$ in the normal state region. 
The dashed and the dashed-dotted lines are the contributions due to
electron-imperfection - and electron-phonon scattering
$W_{e,0}$ and $W_{e,ph}$, respectively.}
\label{fig2}
\end{figure}

Equation \ref{eq2} allows to detemine 
$W_{e,0}$ and $W_{e,ph}$. Shown in figure
\ref{fig2} is the electronic thermal resistivity
$W_e$ of $\rm MgB_2$ displayed 
in the normal state region up to about 80~K.
The solid line is a least squares fit of the data according
to equation \ref{eq2} and the dashed and the dashed-dotted
lines represent $W_{e,0}$ and $W_{e,ph}$, respectively.
Thus, the deduced parameters are $\alpha = 0.55~\rm cm K^2 /mW$ and 
$\beta = 2.8 \times 10^{-7}~\rm cm/mW K$. 
Obviously  from figure \ref{fig2}, the scattering
of electrons with static imperfections of the crystal 
becomes dominant as the temperature approaches $T_c$.

The relative weight  of $\lambda_e$ and $\lambda_l$ at the 
SC transition temperature $T_c$,
as defined from the Wiedemann Franz law,  
also serves to determine the temperature dependence 
of $\lambda_e$ and subsequently of $\lambda_l$
below $T_c$. Since within the BCS theory,
Cooper   pairs do not carry heat and entropy,
the scattering terms of equation \ref{eq2} have 
to be modified in order to account for the decreasing number
of unpaired electrons. 

In the SC state, the thermal
resistivity $W_e^s$ can be represented as
\begin{equation}
W_e^s \equiv 1/ \lambda_e^s = W_{e,0}^s + W_{e,ph}^s
= \frac{\alpha}{T f(t)} + \frac{\beta T^2}{g(t)} ,
\label{eq3}
\end{equation}
with $t = T/T_c$ \cite{Geilikman}.  The functions
$f(t)$ and  $g(t)$ were calculated repeatedly and 
agree well with experimental findings \cite{Bardeen,Tewordt,Geilikman}. 
In the dirty limit of a superconductor, the first
term of equation \ref{eq3} dominates,
i.e., $1/\lambda_e^s \equiv W_e^s \approx W_{e,0}^s$. In terms of the 
BRT theory \cite{Bardeen}  $\lambda_e^s(t)/\lambda_e^n$
is a universal function of $t$, dependent  on the 
value - and the temperature dependence of the 
SC gap $\Delta$.

On the contrary, clean limit superconductors are 
dominated by the second term of equation \ref{eq3},
revealing $1/\lambda_e^s \equiv W_e^s \approx W_{e,ph}^s$. 
Geilikman et al. \cite{Geilikman} have calcutated and tabulated
$g(t)$  which yields again a universal 
behaviour on $t$. Differently to scattering on imperfections,
$\lambda_e^s(t)/\lambda_e^n(T=T_c)$ 
in the BCS limit  increases initially 
with decreasing values of $t$ in spite of a rapid decrease
of electronic excitations. A maximum occurs 
at $t \approx 0.28$ with 
$\lambda_e^s(t=0.28)/\lambda_e^n(T=T_c) = 2.44$. Various high 
temperature superconductors are found to exhibit 
a maximum in $\lambda(T)$ below $T_c$ and thus 
the origin of that feature is, at least partly, 
attributed to a significant scattering strength
of electrons on thermally excited lattice vibrations
(compare e.g. \cite{Castellazzi}).

\begin{figure}[h]
\includegraphics[width=14cm,height=10cm]{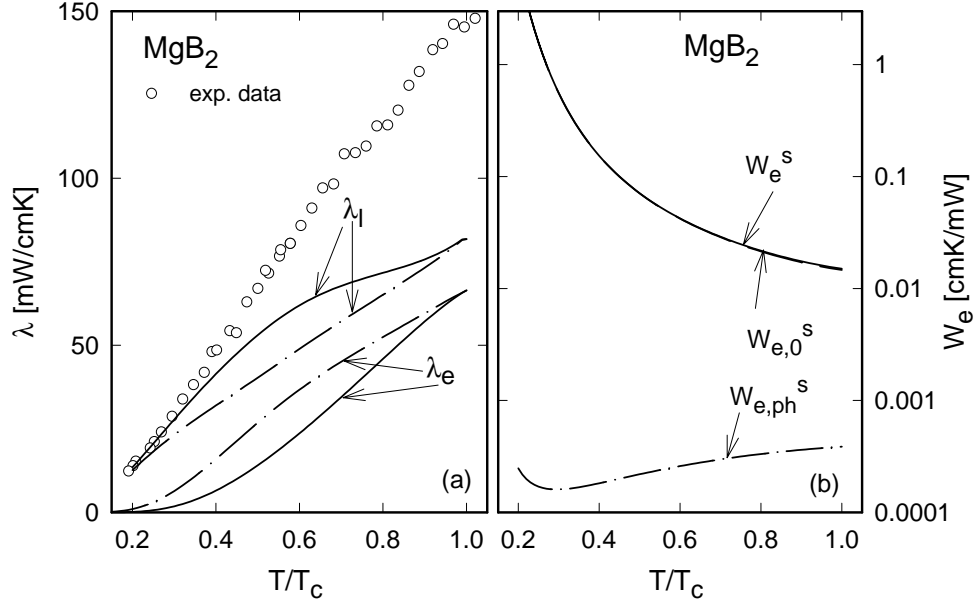}
\caption{(a): Temperature dependent thermal conductivity $\lambda (T)$ 
of $\rm MgB_2$ in the SC state. 
The solid and the dashed-dotted lines represent
electron- and lattice contributions 
$\lambda_e^s$ and $\lambda_l^s$, derived from the 
BCS - and the modified model as discussed in the text, respectively.
(b): Temperature dependent electronic 
thermal resistivity, $W_e^s (T)$ 
of $\rm MgB_2$ in the SC state derived from the BCS model. 
The dashed and the dashed-dotted lines are the contributions due to
electron-imperfection - and electron-phonon scattering
$W_{e,0}^s$ and $W_{e,ph}^s$, respectively.}
\label{fig3}
\end{figure}

Figure \ref{fig2} evidences that slighty above $T_c$,
$W_{e,0}$ exceeds $W_{e,ph}$ by more than one order of magnitude
and contributes at this temperature about 95~\% to
$W_{e}$. This
implicitely favours a description of the thermal conductivity
of $\rm MgB_2$ based on scattering of electrons on impurities. Nevertheless,
for the present analysis of the data
both terms of equation \ref{eq3} are considered in order
to analyse the electronic contribution to the total thermal conductivity.
Taking into account the functions $f(t)$ and $g(t)$
and the numerical values of $\alpha$ and $\beta$
allows us to determine $\lambda_e^s$ below $T_c$
(compare figure \ref{fig3}, panel (a)). $\lambda_e^s$ decreases
with decreasing temperature and is primarily determined by the 
BRT function $f(t)$. The difference $\lambda - \lambda_e^s$ represents
the phonon-originated thermal conductivity $\lambda_l^s$ in the 
SC state. The latter appears to be larger in the SC state
than  $\lambda_e^s$. 
For temperatures $T/T_c < 0.4$, the lattice thermal conductivity 
of $\rm MgB_2$ becomes dominant. This nicely agrees with 
theoretical considerations \cite{Geilikman}.
The lattice term $\lambda_l^s$
is constrained by various scattering processes;
among them are interactions  of the  phonons with electrons,
point defects, dislocations or sheetlike faults.  

To account for observed deviations of the
SC gap of $\rm MgB_2$ from the BCS theory,
the experimental data from tunnel experiments
on  $\rm MgB_2/Ag$ and $\rm MgB_2/In$ \cite{plecenik} have been
used to modify the function $f(t)$ and thus $W_{e,0}^s$. The smaller
the value of $\Delta(T)$ with respect to the BCS theory, the less
steep is the decrease of $f(t)$ when the temperature
is lowered. Since $\Delta (0)$ of $\rm MgB_2$ as obtained from 
that study ($2 \Delta (0) / k_B T_c \approx 2.4$, \cite{plecenik})
is well below the BCS value ($2 \Delta (0) / k_B T_c = 3.5$), 
$\lambda_e^s$ - in this type of analysis - 
becomes larger in the SC state down to $T/T_c \approx 0.2$.
Still, $\lambda_l^s > \lambda_e^s$.
Taking $\lambda_l^s$ as derived from the BCS-like gap, however,  
provides a slightly smoother crossover from the SC -  
to the normal state region of $\rm MgB_2$.

Panel (b) of figure \ref{fig3}
shows the temperature dependent 
thermal resistivity $W_{e}^s \equiv 1/\lambda_e^s$.
Here, the data derived from the BCS model are used.
Obviously, scattering of electrons by phonons 
in the SC state contributes 
just a fraction  to the thermal resistivity $W_{e}^s$, and therefore
$W_{e,0}^s$ is the most significant term below $T_c$.  This, of course,
will not change if the actual dependence of $\Delta (T)$ is considered.

\section{Summary}

The measurement of the thermal conductivity of
$\rm MgB_2$ reveals no pronounced anomaly at $T = T_c$ and 
furthermore no local maximum occurs at temperatures
well below $T_c$. Such a behaviour is most likely caused by
the dominance of electron (hole) scattering on static imperfections
present in the investigated $\rm MgB_2$ sample. In fact, an analysis
based on the BRT model and the model of Geilikman evidences that 
$W_{e,0}^s$ is the predominant term. Beside the classical 
BCS behaviour of the SC gap, an attempt was made to
incorporate the actual gap behaviour reported for $\rm MgB_2$
\cite{plecenik}. 
The subsequent analysis shows that the electronic contribution to the 
thermal conductivity becomes larger than in the former case.
The application of the Wiedemann Franz law to the experimental 
data indicates that both the electronic - and the lattice 
contribution to the total thermal conductivity are almost of
the same size over a large temperature range.

\section{Acknowledgments}

This work is supported by the Austrian FWF P12899.

\newpage

\section{Figure captions}

\end{document}